\documentclass[pdflatex,sn-nature]{sn-jnl}

\usepackage{graphicx}%
\usepackage{multirow}%
\usepackage{amsmath,amssymb,amsfonts}%
\usepackage{amsthm}%
\usepackage{mathrsfs}%
\usepackage[title]{appendix}%
\usepackage{xcolor}%
\usepackage{textcomp}%
\usepackage{manyfoot}%
\usepackage{booktabs}%
\usepackage{algorithm}%
\usepackage{algorithmicx}%
\usepackage{algpseudocode}%
\usepackage{listings}%

\begin{document}

\title[Article Title]{Cosmic dust fertilization of glacial prebiotic chemistry on early Earth}


\author*[1,2]{\fnm{Craig R.} \sur{Walton}}\email{craig.walton@erdw.ethz.ch}

\author[2]{\fnm{Jessica K.} \sur{Rigley}}\email{jessica.rigley@cantab.net}

\author[3]{\fnm{Alexander} \sur{Lipp}}

\author[4]{\fnm{Robert} \sur{Law}}

\author[5]{\fnm{Martin D.} \sur{Suttle}}

\author[1]{\fnm{Maria} \sur{Schönbächler}}

\author[2]{\fnm{Mark} \sur{Wyatt}}

\author[2,6]{\fnm{Oliver} \sur{Shorttle}}

\affil*[1]{\orgdiv{Department of Earth Sciences}, \orgname{Institute für Geochemie und Petrologie, ETH Zürich}, \orgaddress{\street{NW D 81.2, Clausiusstrasse 25}, \city{Zürich}, \postcode{8092}, \country{Switzerland}}}

\affil[2]{\orgdiv{Institute of Astronomy}, \orgname{University of Cambridge}, \orgaddress{\street{Madingley Road}, \city{Cambridge}, \postcode{CB3 OHA}, \country{United Kingdom}}}

\affil[3]{\orgdiv{Department of Earth Sciences, Merton College}, \orgname{University of Oxford}, \orgaddress{\city{Oxford}, \postcode{OX1 4JD}, \country{United Kingdom}}}

\affil[4]{\orgdiv{Department of Earth Sciences}, \orgname{University of Bergen}, \orgaddress{\city{Bergen}, \postcode{NO-5020}, \country{Norway}}}

\affil[5]{\orgdiv{School of Physical Sciences}, \orgname{Open University}, \orgaddress{\street{Walton Hall},\city{Milton Keynes}, \postcode{MK7 6AA}, \country{United Kingdom}}}

\affil[6]{\orgdiv{Department of Earth Sciences}, \orgname{University of Cambridge}, \orgaddress{\street{Downing Street},\city{Cambridge}, \postcode{CB2 3EQ}, \country{United Kingdom}}}


\abstract{Earth's surface is deficient in available forms of many elements considered limiting
for prebiotic chemistry. In contrast, many extraterrestrial rocky objects are rich in these same elements. Limiting prebiotic ingredients may therefore have been delivered by exogenous material; however, the mechanisms by which exogeneous material may be reliably and non-destructively supplied to a planetary surface remains unclear. Today, the flux of extraterrestrial matter to Earth is dominated by fine-grained cosmic dust. Whilst this material is rarely discussed in a prebiotic context due to its delivery over a large surface area, concentrated cosmic dust deposits are known to form on Earth today due to the action of sedimentary processes. Here, we combine empirical constraints on dust sedimentation with dynamical simulations of dust formation and planetary accretion to show that localised sedimentary deposits of cosmic dust could have accumulated in arid environments on early Earth -- in particular glacial settings that today produce cryoconite sediments. Our results challenge the widely held assumption that cosmic dust is incapable of fertilising prebiotic chemistry. Cosmic dust deposits may have plausibly formed on early Earth and acted to fertilise  prebiotic chemistry.}

\keywords{origin, life, cosmic, dust}



\maketitle

\section{Introduction}\label{sec1}

The origin of life on Earth probably resulted from interacting solid, liquid, and gaseous reservoirs of bioessential elements in reactive molecular forms \cite{Sasselov:2020}. Experimental work demonstrates that high to moderate concentrations (mM to 100's mM) of simple species (e.g., HCN, $\mathrm{PO_4^{3-}}$, and $\mathrm{HSO_3^{-}}$) can produce high yields of biologically-relevant molecules such as nucleic acids, lipids, and peptides \cite{Powner:2009} \cite{Rimmer:2018} \cite{Xu:2020} \cite{Bonfio:2020} \cite{Foden:2020}. However, a remaining gap in understanding the geological context of prebiotic chemistry on Earth is the mechanisms by which concentrated feedstocks were produced \cite{Graaf:2000}.

Common terrestrial rocks are relatively poor in reactive and/or soluble forms of the key elements mentioned above: phosphorus (P), sulfur (S), nitrogen (N), and carbon (C). Indeed, life on Earth is engaged in fierce competition for the limited endogenous bioavailable reservoirs of these elements. Complex enzymatic machinery has evolved in response to this challenge, such that life can extract these species from the environment even when they occur in limited concentration or in largely inert chemical form. The pre-enzymatic world of prebiotic chemistry must have initially lacked such mechanisms to enhance the availability of key species. However, certain processes in Earth's early history may have gone part or all of the way towards solving this apparent paradox. One such possibility is the accretion and surficial sedimentary sorting of cosmic dust (here defined as grains of size $\mathrm{<3\;mm}$).

Cosmic dust is comprised of the mineral grain aggregates produced by collisions between asteroids \cite{Nesvorný:10.1038/nature00789} and/or the sublimation and disintegration of comets \cite{Rigley:2020} (Fig. 1a). Such particles produced further from the Sun can then drift inwards due to Poynting-Robertson (P-R) drag and be accreted by Earth. Cosmic dust contains bioessential elements (for terrestrial life), e.g., P, S, N, and C, at concentrations well above that of Earth’s crust \cite{Carrillo:2020, Rojas:2021, Rudraswami:2021}. {Many cosmic dust grains represent nearly pristine samples of their parent body objects, which appear to span comets (volatile-rich outer Solar System materials) and asteroids (a mixture of early-formed differentiated objects and relatively late-formed and/or small-to-moderate sized undifferentated objects, of both outer and inner Solar System origin) }\cite{Suttle:2020, Rojas:2021}.

{In contrast to larger objects, the flux of cosmic dust to Earth is essentially constant on yearly timescales. Moreover, some fraction of cosmic dust grains pass relatively gently through the Earth's atmosphere, thereby retaining a greater fraction of primitive CHNS than do large (e.g., bolide-type) impactors} \cite{Matrajt:2006, Furi:2013, Carrillo:2015, Mehta:2018}. {The continuous and mostly non-destructive accretion of cosmic dust across planetary surfaces suggests a significant advantage over discrete and violent episodes of bioessential element delivery by larger impactors, potentially greatly improving the chances of the successful entry of extraterrestrial matter into prebiotic chemistry.}

Cosmic dust is accreted across the Earth's entire surface area, such that it is initially dilute in terms of mass per unit area relative to discrete delivery events by larger objects \cite{Pearce:2017bab}. On this basis, the prebiotic importance of cosmic dust has previously been questioned \cite{Pasek:2008, Pearce:2017bab}. However, there are many planetary processes that can concentrate fine grained materials from across large surface areas to form concentrated deposits, e.g., aeolian, fluvial, and glaciogenic sorting mechanisms, which produce dunes, beaches, and moraines, respectively. Indeed, these mechanisms are known to operate today to locally concentrate cosmic dust by up to 1000-fold relative to the global baseline; on the basis of which, cosmic dust has been proposed to be relevant to the origin of life \cite{Maurette:1986, Maurette:1995, Rochette:2008, Genge:2018, Longo:2018, Tomkins:2019}. {Despite these possible links, there are no existing quantitative models of the supply to and surface cycling of cosmic dust on early Earth.}

Here, we use astrophysical simulations and geological models to quantify both the flux and compositions of cosmic dust accreted to the surface of early Earth (Fig. 1a). We combine these results with geological models of subaerial concentration in a range of environments (Fig. 1b-c). Using our results, we assess the possible utility of cosmic dust deposits on early Earth for fertilising prebiotic chemistry.

\section{Results}\label{sec2}

\textbf{Estimating early dust accretion fluxes}

We simulate cosmic dust accretion to Earth during the first 500 million years after the Moon-forming impact at 4.51 Ga \cite{Barboni:2017}. {Numerical models (see Methods) were used to simulate the generation, interplanetary transport, and terrestrial accretion flux of dust produced by two compositionally distinct parent body populations: Jupiter-family comets (JFCs) and asteroids (Fig. 1). Asteroidal dust is considered as the sum of dust generated by Main Belt asteroids and unstable planetesimals located in planet-forming regions (see Methods).}

Comparing the results of our numerical model to estimates of present day cosmic dust accretion flux to Earth, we find that total early cosmic dust accretion fluxes would have been on the order of {100-to-10,000-fold higher than observed today, depending on the value chosen to represent the modern flux (Fig. 2--3)} \cite{Plane:2013}. These elevated dust fluxes arise mainly from (i) a more massive main asteroid belt leading to higher rates of collisional grinding \cite{Davison:10.1111/maps.12193}, (ii) dynamical perturbation of the asteroid belt and injection of cometary objects by giant planet migration \cite{Rigley:2020}, and (iii) the collisional erosion of the population of rocky objects left over from planet formation, which had yet to be depleted.

Our simulations suggest that early cosmic dust fluxes would have been dominated by fragments of the dynamically unstable asteroid population and comets, with a relatively minor contribution by the Main Belt (Supplementary Fig. 3). Short-lived spikes in the proportion of cometary material in the cosmic dust accretion flux occur as a result of the scattering of large or long-lived comets, with these spikes lasting for on the order of one to several Myr. The timing of these cometary dust spikes has no special sigificance in the models, simply reflecting a particular realisation of a semi-random history. Critically, this means that intervals of high-flux comet-dominated dust supply could have occurred at any point during the first 500 Myr of Earth history.

Grain size frequency distributions for the mean mass accretion flux of all three cosmic dust sources are bimodal between 1 and 3000 $\mathrm{\mu m}$, peaking at around 1--10 and 100--500 $\mathrm{\mu m}$ (Supplementary Fig. 1). These distributions are ultimately very similar to those arriving at Earth's atmosphere and observed in cosmic dust deposits today \cite{Love:1991, Suttle:2020, Carrillo:2020}, with a somewhat larger contribution by particles larger than 500 $\mathrm{\mu m}$. {Overall, our model predicts an early cosmic dust flux that differs mainly in having a higher total mass per unit time and distinct compositional make-up compared with that observed today.} However, like today \cite{Nesvorný:2010}, {and with great relevance for prebiotic chemistry, volatile-rich dust grains -- including $>\;$65$\%$ cometary material at some points -- are expected to have dominated the cosmic dust flux to early Earth (Fig. 2b).}

\textbf{Occurrence and composition of cosmic dust deposits}

The final concentration of cosmic dust in a sedimentary deposit will depend on dust input rate, endogenous sediment supply, and local dust concentration mechanisms. Using our estimates of dust accretion fluxes to prebiotic Earth, we construct a model to predict the proportion of dust within coeval unconsolidated sediments (see Methods; Fig. 1c). We consider a range of end-member sedimentary environments: glacier surfaces, hot deserts, and deep-sea sediments. The proportion of cosmic dust by mass, the relative proportions of types of cosmic dust, and their chemical compositions and degrees of alteration have been measured for all of these environments on Earth today, allowing us to ground truth our calculations (Supplementary Table 1). We use the resulting empirically-derived scaling factors to estimate both the total cosmic dust concentration by mass and relative abundance of cosmic dust-types within sediments on early Earth as a function of early dust flux relative to modern (Fig. 3).

Many aspects of early Earth remain uncertain. To obtain meaningful results, we assume that all early Earth environmental parameters are equivalent to present day except the rate of cosmic dust accretion. This minimises sources of uncertainty that would otherwise be difficult or impossible to constrain, i.e., allowing us to test the prebiotic relevance of dust supply in one well constrained early Earth scenario.

We further assume that the concentration mechanisms operating in these environments will operate with the same efficiency at all accreted mass fluxes of dust considered in our work, i.e., sedimentary environments are far from being dust saturated. We focus on environments where high cosmic dust concentrations are achieved without a high degree of fractionation with respect to different categories of dust. This allows us to assume that differences in the compositional make-up of the accreted cosmic dust translates directly into the resulting estimated compositions of putative early Earth cosmic dust deposits \cite{Maurette:1987, Suttle:2020}. Furthermore, although the presence of vegetation in modern-day environments strongly affects sedimentation mechanisms and early Earth would have lacked such interferences, the modelled environments in our study are essentially absent of any vegetation, such that our all-else-being-equal approach is applicable.

{To interpret our results, we must normalise our values for early Earth cosmic dust accretion flux to a reference value for the modern top-of-atmosphere dust accretion flux. We compare to a range of conservative values for the top of-modern-atmosphere dust accretion flux: a high value of 100 t/d and a low value of 10t/d } \cite{Plane:2013}. {Given that our model for predicting the proportion of cosmic dust in early sediments is pegged to present day systems, using lower values to represent the modern dust flux renders our predicted early dust flux relatively higher and hence predicts higher cosmic dust proportions in early sedimentary systems (Fig. 3).}

We find that cosmic dust represents a minor component of deep ocean sediments even at the highest dust accretion rates estimated by our model (Fig. 3). Meanwhile, cosmic dust may have represented $>$ 50 \% of sediments in desert and glacial settings given the same cosmic dust fluxes. The highest concentrations ($>$ 80 \%) would have occurred within glacial ablation zones in cryoconite-type sediments (Fig. 3), just as these sediments contain the highest reported cosmic dust concentrations today \cite{Maurette:1986}.  {Cryoconite forms on Earth today in small (metre-to-centimetre) melt-water pools on the surface of retreating ice-sheets, with dust components non-destructively introduced by wind-driven transport} \cite{Bagshaw:2013}. Compositionally, cosmic-dust-rich analogues of these sedimentary environments would represent a unique and compelling environment for fostering prebiotic chemistry. 

{We can gain an insight into this plausible prebiotic utility by exploring the likely chemical compositions of putative cosmic sediments. This is achieved by combing estimates of the contributions of cometary versus asteroidal dust with estimates of their respective average C, N, P, and S concentrations (Supplementary Table 1). Whilst other arid environments are plausible settings for concentrated deposits of cosmic dust, e.g., hot deserts (Fig. 3), we focus here on glaciogenic cosmic sediments -- the environment we expect to be richest in cosmic dust for any given dust accretion flux (Fig. 3--4). Moreover, glacial environments are unusual in being generally arid yet always having the capacity to generate liquid water, unlike hot arid environments. Our results show that continuous accretion of cosmic dust and concentration in glaciogenic settings could have generated sediments enrichmented in bioessential elements relative to average Upper Continental Crust} \cite{Rudnick:2014} {by up to 100-fold.}

{These enrichments peak for S, N, and C during episodes of enhanced cometary delivery. The highest amplitude comet-delivery spikes that our model predicts result in around 2-fold higher concentrations of N and C in the cyroconite sediments than achieved during background cosmic dust accretion. The highest possible enrichments may be obtained in the timeframe immediately following the collisional breakup of parent body objects. These episodes of enhanced delivery (million-year timescales) may outpace background cosmic dust fluxes by around 1 order of magnitude} \cite{Schmitz:10.1126/sciadv.aax4184} {(Fig. 3--4), yielding equivalently enriched sediment compositions.} {Given the rarity of parent body break up events over the last 4 billion years, ground truth constraints on the impact of post-break-up dust fluxes on concomitant proportions by mass of cosmic dust in terrestrial sediments are hard to establish. However, our estimate of peak fluxes of around 10,000x modern, lasting on the order of millions of years, is supported by observations of cosmic dust, micrometeorite, and fossil meteorite abundance in Ordovician sediments }\cite{Heck:10.1038/s41550-016-0035, Schmitz:10.1126/sciadv.aax4184}, {which record the aftermath of the L chondrite parent body disruption event at 474 $\pm$ 22 Ma} \cite{Walton:2023}.





Our results highlight that cosmic dust-rich deposits would have formed most readily in environments of low endogenous sediment production on early Earth over much of its early history, with exceptionally high cosmic dust concentrations developing in the aftermath of individual parent body break-up events. In particular, we find that cosmic dust proportions would have been high in glacial settings -- specifically, cryoconite fields within glacial ablation zones (Fig. 3--4). Here, cosmic sediments ($>$ 50 \% cosmic dust by mass fraction) would have formed, enriched with respect to average crustal rocks both in terms of the fraction of bioessential elements in reduced form and the overall concentration of those elements. This outcome is robust, given that early sources of cometary material were chemically comparable to the comets sampled by recent missions \cite{Della:2020} and unmelted carbonaceous micrometeorites \cite{Matrajt:2003, Suttle:2020} (Supplementary Table 1).

\section{Discussion}\label{sec12}

In principle, cosmic dust rich cryoconite is a compelling scenario for prebiotic chemistry (Fig. 5). However, there are also caveats that should be highlighted. A possible challenge for the prebiotic relevance of our scenario is that many ice sheet surfaces undergoing melting are semi-permeable and connected to a wider supraglacial drainage system \cite{Cooper:2018} (Fig. 5). Cryoconite most often forms in the ablation zone of glaciers, where ice dynamics or the continual development of the supraglacial drainage system typically preclude permanent or even multi-annual cryoconite holes. As such, prebiotically interesting species leached from dust-rich deposits in cryoconite holes may drain along with melt water to the level of the water table (Fig. 5a) -- and be diluted in the process \cite{Cooper:2018}.


In contrast, the dry valleys of Antarctica shelter cryoconite holes which form with an ice surface lid and cold surrounding ice limiting lateral water transport. These result from the particular and unique conditions of the dry valleys: extreme aridity and cold, with surface melting possible for only brief periods in summer. Solar radiation heats low-albedo inclusions within the ice which melt the ice around them in a local greenhouse effect, despite the icy lid \cite{Bagshaw:2013} (Fig. 5b). Here, biogeochemical recycling results in the waters of cryoconite holes becoming 100-fold enriched in bioavailable forms of limiting nutrients relative to surrounding ecosystems \cite{Bagshaw:2013}. Abundant life occurs in such systems mostly supported by leached nutrients from the sedimentary dust deposits at the base of the cryoconite holes \cite{Wadham:2019}. In a prebiotic Earth scenario, we may therefore expect similar cosmic-sediment-filled environments to form on the surface of impermeable ice lids, abiotically-enriched in prebiotic feedstock.


Nonetheless, even Dry Valleys type cryoconite deposits are subject to transport through ice dynamics of up to 20 m/yr \cite{Kavanaugh:2009}, precluding long-term formations. However, the regular destabilisation of cryoconite sediments directly supplies meltwater and sedimentary material to endorheic proglacial lakes (Fig. 5c). Indeed, ecosystems in these closed proglacial lakes are known to depend on fertilisation by cryoconite-derived nutrients \cite{Fountain:2004, Bagshaw:2013}. Glacier margins therefore provide settings capable of both locally concentrating cosmic dust and initiating closed system aqueous prebiotic chemistry with the products of cosmic dust dissolution/leaching.

Both cosmic-dust-rich cryoconite sediments and endorheic proglacial lakes would appear to have many attractive properties for both initiating and sustaining prebiotic chemistry. These environments would represent networks of icy prebiotic chemical `reactors', replete with the known advantages of ice-hosted prebiotic chemistry: freeze-thaw wet-dry cycles (annual--daily) \cite{Attwater2010}, low water-rock ratios, potential for UV irradiation as well as UV shielding \cite{Patel:2015, Rimmer:2018}, and the ability to exchange over weekly-to-yearly timescales with other cosmic-dust-filled reactors \cite{Maurette:1986} -- analogous to the stream intersection models envisaged by Sutherland \cite{Sutherland} and Rimmer \cite{Rimmer:2023}. 

The geochemistry of cosmic-dust-rich sediments provides unique advantages over and above glacial settings hosting sediments of solely terrestrial geochemical character. In particular, identifying soluble early Earth sources of P and S is a long running challenge \cite{Schwartz:2006, Ranjan:2018}. Prebiotic chemistry directly initiating after exposure of cosmic dust deposits to liquid water in e.g., cryoconite holes or proglacial lakes would have had access to reactive P- and S-rich materials of a very fine grain size, namely:

\begin{itemize}
    \item P (up to 1500 ppm) -- schreibersite ($\mathrm{Fe_3P}$), apatite ($\mathrm{Ca_5PO_4[OH,Cl,F]}$), merrillite ($\mathrm{Ca_9NaMg(PO_4)_7}$) \cite{Walton:2021b};
    \item S (up to 5 wt\%) -- troilite (FeS), pyrrhotite ($\mathrm{Fe_{1-x}S}$),  pentlandite $\mathrm{(Fe, Ni)_9S_8}$, pyrite $\mathrm{(FeS_2)}$, chalcopyrite $\mathrm{(Cu, Fe)S_2}$, sulfonic acids, \cite{Wood:2019, Naraoka:2023}.
\end{itemize}

These materials are analogous to the powders used to accelerate dissolution of chemical reagents in a laboratory setting -- often with a grain size of tens to hundreds of microns, similar to the size distribution of cosmic dust particles \cite{Maurette:1987, Suttle:2020}. Such fine particles sizes aid with dissolution and hence with sustaining high throughput of key species for prebiotic chemistry. Indeed, glacial meltwater discharge zones are known today to yield among the highest known fluvial fluxes of reactive and particulate phosphate on Earth \cite{Hawkings:2016}.

Some non-trivial fraction of the P content of early cryoconite sediments would have been speciated as phosphide (perhaps around 50\%) \cite{Walton:2021b}. The action of ultra-violet light combined with dissolved $\mathrm{H_2S\;and\;HS^-}$ species has been shown to be capable of oxidising species released during phosphide corrosion in water, forming phosphate \cite{Ritson:2020} -- a key constituent in and catalyst for prebiotic chemistry \cite{Islam}. Sulfide and its derivative oxidised species (e.g., bisulfite) have further been shown to assist the prebiotic syntheses of nucleic acids \cite{Rimmer:2018} and many components of central carbon metabolism \cite{Liu:2021}. Given the apparent scarcity of high throughput sources of fine grained sulfide, phosphate, and phosphide on Earth \cite{Schwartz:2006, Ranjan:2018}, we suggest that cosmic dust sedimentation in glacial systems provides a highly relevant geological feedstock for prebiotic chemistry. 


Our results show that Antarctic-like ice sheets on prebiotic Earth would have plausibly given rise to aerially extensive regions of cryoconite-hosted cosmic dust-rich deposits and derivative proglacial lakes, with many attractive properties for prebiotic chemistry. Our results therefore highlight a particular environmental context for dust-fed scenarios for prebiotic chemistry, linking the relevance of the scenario to the climate of early Earth. This provides a helpful line of reasoning by which our scenario can be falsified on planetary climatic grounds alone. At present, it is unclear whether or not the glacial environments needed to forge concentrated cosmic dust deposits would have been common on early Earth. Glaciers are known to have formed up to 2.5 billion years ago on Earth \cite{Kirschvink:2000}. Recent models also suggest a cold early Earth \cite{Kadoya:2020}, coincident with the timeframe explored in our study and which could support our scenario. 


With regard to C and N, we suggest that cosmic dust may also have been a stockpiling agent for prebiotic chemistry, rather than a direct participant. Alongside the arrival of single large impactors \cite{Todd:2020}, cosmic dust sedimentary cycling may be a plausible mechanism by which to involve C- and N-rich cometary matter in prebiotic chemistry. Whilst the soluble component of cosmic dust will be liberated rapidly during interaction with liquid water, much of the C and N content of cosmic dust arrives in the form of mostly insoluble and inert species. These include N-bearing kerogen (largely poly-HCN), N-heterocycles, amino acids, amines, amides, purines, polar species, aromatics, hydroxy, dicarboxylic, and carboxylic acids \cite{Pizzarelo:2010}.

Instead of suggesting irrelevance to prebiotic chemistry, the inert character of many of these species strongly implies their capacity to accumulate over time. In settings with low rates of terrestrial sediment deposition -- glacier surfaces, for example -- cosmic dust could have built up sedimentary blankets with around 0.2 wt\% N and 5 wt\% C (Fig. 4c). Transport fractionation of organics relative to silicate phases could have generated still greater C- and N-enrichment. A pure nitrogenous kerogen sedimentary end-member would have around 3 wt\% N and 70 wt\% C \cite{Lawler:1992}. Indeed, extreme organic C enrichments are observed in cryoconite sediments with increasing distance inland across the Greenland ice sheet albation zone -- at least in part due to transport fractionation of organic materials \cite{Stibal:2010}. A key caveat is that such sediment compositions are only possible assuming that atmospheric entry of dust through a reducing anoxic atmosphere promotes minimal loss of N, in contrast to the substantial (90--99\%) loss observed during entry through Earth's present day oxidising atmosphere (Fig. 4c) \cite{Marty:2005}.

Stockpiling mechanisms are commonly invoked in proposed scenarios for cyanosulfidic prebiotic chemistry \cite{Sutherland}. The thermal processing e.g., meteorite impacts \cite{Sutherland} or melting into magma chambers \cite{Bird:1981} of inert CN-rich and relatively H-poor reducing sediments could have supplied prebiotically relevant concentrations of HCN to overlying aqueous environments \cite{Rimmer:2019}. Such a mechanism of continuous HCN supply could be made possible by an intermediate stage of sedimentary sorting and stockpiling of dust-derived inert CN-bearing organics. Via a cosmic dust sedimentation and heating pathway -- closely analogous to volcanic remobilisation of sedimentary organic C on Earth today \cite{Rimmer:2019} -- the requisite starting ingredients for cyanosulfidic prebiotic chemistry could plausibly have been delivered, concentrated, and liberated to clement environments for the origin of life.

We have discussed different geological mechanisms to separately supply P + S and C + N from cosmic dust to prebiotic chemistry. However, whilst these mechanisms operate across different spatial scales and timescales, they are not mutually exclusive. Thermally processed deposits of exogenous organic matter may have degassed HCN into lake systems fed by glacial meltwater rich in P and S sourced from locally dissolving cosmic dust. Taken together with recent findings from geology, astronomy, and prebiotic chemistry, our results provide support for the fertilisation of prebiotic chemistry by cosmic dust on early Earth. Furthermore, cosmic dust is potentially a widespread and flexible planetary fertiliser, being accreted in quantities that may be assessed by observation \cite{Kral:2018, Rigley:2020} to potentially habitable exoplanets.

\section{Methods}

\subsection{Cosmic dust deposit model parameters}

Cosmic dust grains in pristine modern sedimentary systems appear to derive mainly from undifferentiated carbonaceous-chondrite- and/or cometary precursor bodies \cite{Taylor:2011, Goderis:2020}. Cosmic dust that appear to originate from differentiated asteroids represent less than 5 percent by mass of all cosmic dust in most deposits \cite{Soens:2022}. The compositional disparity between modern cosmic dust accretion fluxes and predicted early fluxes must be considered in interpreting the significance of our results for supplying bioessential elements to early Earth's surface. Bridging this gap requires us to make assumptions both about the parent bodies of cosmic dust falling to Earth today and how those precursor objects are represented in our simulations.

We assume that all cometary material is well represented by the compositions of unmelted micrometeorites and apparently cogenetic carbonaceous chondrite meteorites \cite{Soens:2022}. Dust sourced from asteroids is conservatively assumed to sample differentiated precursor bodies. Cosmic dust grains sourced from differentiated material is considered to represent an equal mixture of crust- and core- material, and to be well represented by the compositions of silicate- and iron-type grains arriving on Earth, today \cite{Badjukov:2010}. These assumptions inform the values in Supplementary Table 1.

Particle diameter exerts a strong control on the survival of cosmic dust during atmospheric entry (Supplementary Fig. 1). The majority of particles with sizes $\mathrm{<\;0.1\;mm}$ survive to the Earth's surface with minimal heating \cite{Love:1991} and empirical studies demonstrate they retain a significant proportion of their temperature-sensitive organic matter \cite{Furi:2013}. We estimate the proportion of early accreted material that survives unaltered versus that which melts and that which vaporises using the results of Love and Brownlee \cite{Love:1991} (Supplementary Fig. 1). Again, we take an all-else-being-equal approach to justify this approach, where the atmosphere of early Earth is considered to be of similar mass to today.

Our findings are robust to uncertainties regarding the assumed composition of the terrestrial sedimentary component. Whilst we plot all results relative to and assuming a modern day average Upper Continental Crust for terrestrial sediment within the modelled early dust deposits, different choices would result in higher reduced fractions and/or nutrient enrichment factors. This is because estimated early crustal compositions that differ from modern average Upper Continental Crust are instead more similar to Earth's mantle \cite{Rudnick:2014} -- thereby containing lower concentrations of P, S, C, and N, and resulting in their stronger relative enrichment in early cosmic dust-rich sedimentary deposits. We have not modelled the leaching of bioessential elements from dust during sedimentary transport. This assumption is valid for the arid conditions and aeolian transport mechanisms operating in the desert and glacial settings that favour cosmic dust deposit formation.

Whilst stochastic and relatively short-lived on geological timescales (several Myr), intervals of comet-dominated dust supply occur with high probability in any given simulation run. These intervals are also sufficiently long-lived to allow complete cycling of cosmic dust through all of the surface environments that we consider \cite{Genge:2018, Suttle:2020}. Therefore, whilst background dust supply provides notable enrichments, our results highlight that stochastic large cometary breakup events are crucial for producing dust-rich sediments with the highest concentrations of bioessential elements.

\subsection{Estimating cosmic dust abundance within early sediments}

We take an all-else-being-equal approach to assess the potential for cosmic sediment formation on early Earth. Critically, we assume that the deposition flux of dust in a given environment directly scales with the flux accreted to Earth, i.e., that processes acting to concentrate dust operate with equal efficiency at higher dust accretion rates. This assumption is defensible because maximum sediment loading is expected to be extremely low in the low sedimentation environments that we consider, even given much higher early cosmic dust accretion fluxes.

Given these assumptions, we can set up a linear relationship between the cosmic dust accretion flux to early Earth relative to the at-atmosphere flux estimated today ($\epsilon_{x}$) and the ratio of cosmic dust mass per unit mass of sediment in a given sedimentary system ($\delta$, $\mathrm{kg/kg}$). As cosmic dust fluxes increase, total sediment within the system of interest will increase. To track this, we must include as a constant the original mass of terrestrial sediment mass per unit mass of sediment $(1-\delta)$ in the sedimentary system. The fraction of cosmic dust in the putative early sedimentary system ($f_{dust}$) is then given by 






\begin{equation}
    f_{dust} =\frac{\delta_{modern}\;\epsilon_{x}}{\delta_{modern}\;\epsilon_{x}+(1-\delta_{modern})}.
    \label{eq:1}
\end{equation}




The principal parent body sources are considered to be asteroids and comets,

\begin{equation}
    f_{dust} = \frac{\sum_{x=cometary,asteroidal}F_{x}\;\delta_{modern}\;\epsilon_{x}}{\sum_{x=cometary,asteroidal}F_{x}\;\delta_{modern}\;\epsilon_{x} +(1-\delta_{modern})}.
\end{equation}






We can calculate the concentration of a given element in the early sedimentary deposit ($\gamma$) as as mixture of the chemical compositions ($\beta$) (Supplementary Table 1) \cite{Wasson:1988, Schramm:1990, Javoy:1997, Matrajt:2003, Marty:2005, Furi:2013, Rudnick:2014, Prasad:2018, Tomkins:2019, Suttle:2020} of endogenous sediment (E) and cosmic dust as follows,

\begin{equation}
    \gamma = \frac{\sum_{x=cometary,asteroidal}F_{x}\;\delta_{modern}\;\epsilon_{x}\;\beta_x + (1-\delta_{modern})\;\beta_E}{\sum_{x=cometary,asteroidal}F_{x}\;\delta_{modern}\;\epsilon_{x}\ +(1-\delta_{modern})}.
\end{equation}

The proportion of the early dust sedimentation flux made up by each cosmic dust class will have also been partially determined by the effects of atmospheric entry. Volatiles may also be preferentially lost during atmospheric entry heating of dust grains \cite{Matrajt:2003, Matrajt:2006, Furi:2013}. We use empirical observations of the volatile chemistry of cosmic dust particles to inform our calculations, which show that S, C, and P are largely conserved during entry, even if they may be to some extent chemically reorganised \cite{Riebe:2020}. In contrast, a large fraction of nitrogen is lost from cosmic dust particles during atmospheric entry, largely due to degradation by oxidation \cite{Marty:2005}. Entry into an anoxic and possibly reducing early atmosphere, relevant for early Earth, may have been far less efficient for destabilising and removing nitrogen from dust grains -- an end-member scenario for which we consider here.

Since cosmic dust vaporisation during atmospheric entry is heavily dependent on particle size \cite{Love:1991}, we examined whether the size frequency distribution of early dust accretion fluxes predicted by our model differs to that observed today. We use the results of Love and Brownlee \cite{Love:1991} to estimate the proportion of all cosmic dust that is vaporized versus melted/survived unaltered during atmospheric entry today. In order to facilitate like-for-like comparison between our results and those of Love and Brownlee \cite{Love:1991}, we generate continuous distributions of accreted mass as a function of grain size from histograms of mean mass accreted per grain size bin. We interpolate between the centre point of each bin. These continuous distributions are then used to predict the fraction of early cosmic dust grains that vaporize ($f_v$) during entry for each source population (Supplementary Fig. 1).

{Overall, the size distributions of material arriving at the top of the atmosphere and arriving at the surface after entry are essentially similar to those observed today} \cite{Suttle:2020}. {However, the exact abalation behaviour may differ for each population based on mineralogical and entry velocity differences} \cite{Carrillo:2015, Carrillo:2020}. {To largely eliminate this uncertainty, we proceed with our approach of anchoring to empirical observations. We compare the flux estimate of cometary versus asteroidal dust arrivin at Earth's surface to top of atmosphere flux estimates to obtain an empirical value for }($f_v$). {Similarly, we use empirical observations of unmelted to melted dust particles to estimate the fraction that experiences melting during entry }($f_m$),{ and compositional analyses to estimate the fraction of volatiles lost during entry from both unmelted and melted particles} ($f_{loss}$). This approach then yields,




\begin{equation}
    \gamma = \frac{\sum_{x=cometary,asteroidal}\;(1-f_{v})\;(\theta_{x}\;\beta_{x}\;f_m\;f_{loss} + \;\theta_{x}\;\beta_{x}\;(1-f_m))\;+\;(1-\delta_{modern})\;\beta_E}{\sum_{x=cometary,asteroidal}\;(1-f_{v})\;(\theta_{x}\;\beta_{x}\;f_m\;f_{loss} + \;\theta_{x}\;\beta_{x}\;(1-f_m)) + (1-\delta_{modern})},
\end{equation}

where $\theta_x$ is equivalent to $F_{x}\;\delta_{modern}\;\epsilon_{x}$.



Uncertainty on the measured masses of cosmic dust per kg of sediment is obtained from literature reports. These literature reports were conducted at different times, in different places, and using different methods to both obtain sediment and to filter cosmic dust from the material. Ultimately, uncertainty on the proportion of cosmic dust in the sediment is trivial compared to that introduced by assumptions made during the construction of the dust flux model. For example, rounding up reported 2$\;\sigma$ uncertainty on cosmic sediment fraction to the nearest interval, uncertainties of around 10 $\%$  characterise the majority of measurements. Details of the values of all parameters used in our model for cosmic sediment composition can be found in Supplementary Table 1.




\subsection{Numerical model of collisional dust generation and transport to Earth}

\medskip

{We simulated the distributions of dust produced by two compositionally distinct parent body populations: Jupiter-family comets (JFCs) and asteroids. Asteroids were treated as the sum of two dynamically distinct populations: relatively stable Main Belt objects and rapidly depleted unstable planetesimals leftover from planet formation.} For each population, the overlap of the dust grain orbits with Earth was used to calculate the resulting fluxes of cosmic dust accreted onto Earth. 

Intrinsic to this model are assumptions about the history of the Solar System and its formation. For the start time of our models, we use the Moon-forming impact. Models in the literature suggest that this occurred early, $\sim 50-100$~Myr after the formation of CAIs \cite{Kleine09,Jacobson14,Bottke15}. We assume the Moon-forming impact occurred at 50~Myr, which is time zero of our simulations. We also assume that the giant planet instability occurred early \cite{Clement18,Mojzsis19}. Simulations were run for 500~Myr for each source population. 

We modelled the distributions of dust produced by each source population separately using a kinetic model \cite{vLieshout14} which follows the evolution of a population of particles in terms of their orbital elements and sizes. The numbers of particles in bins of particle size, pericentre, and eccentricity evolve due to several forces. Destructive collisions between particles remove particles that collide and produce smaller fragments; P-R drag causes particles to lose angular momentum and drift towards the Sun; and radiation pressure acts radially outwards on small particles.

\textbf{Comets}

The dust produced by JFCs was simulated using the model of \cite{Rigley22}, which was originally created to model the production of the present zodiacal cloud via spontaneous fragmentation of comets. We updated the model to allow comets to be scattered in at a variable rate. In the early Solar system, comets should be scattered inwards at a much higher rate than today. The rate of comet scattering was extrapolated from N-body simulations which followed the distribution of JFCs throughout the 4.5~Gyr history of the Solar system \cite{Nesvorny17}. The input rate of comets was found to be tens to hundreds of times higher in the first 500~Myr of the Solar system than the present rate, declining with time (Supplementary Fig. 2). This cometary model gives the distribution of dust produced by comets with time, which was then input into the kinetic model to find how the dust evolves after it is released from the comets. 

\textbf{Asteroids}

For the asteroid belt, the kinetic model was initialised with bodies with the size distribution \cite{Bottke05} of the current belt, going from diameters of 1 to 1000 km. The initial orbits of asteroids were assumed to be those of the present main belt (JPL Small-Body Database \newline (https://ssd.jpl.nasa.gov/tools/sbdb\_query.html). The initial mass of the asteroid belt and its subsequent evolution is poorly constrained. While collisional evolution of asteroids will slowly deplete the overall mass of the belt, it is likely that the dynamical evolution of planets dominated depletion of the asteroids (e.g. \cite{Clement19}). Based on \cite{Morbidelli15}, we assume that following the giant planet instability, the asteroid belt had a mass four times higher than its present mass ($\sim 5 \times 10^{-4}~\mathrm{M}_\oplus$ \cite{Krasinsky02,Kuchynka13}), and lost half of its mass over the subsequent 100 Myr by depletion of unstable resonances. We therefore initialised the model with a mass of $2\times 10^{-3}~\mathrm{M}_\oplus$ of asteroids. The asteroids then evolve in the kinetic model by collisional evolution, producing dust, which further evolves due to collisions and radiation forces. 

\textbf{Rapidly depleted asteroids}

Similarly, the collisional evolution of asteroids leftover from terrestrial planet formation was traced with the kinetic model. The size distribution was assumed to be the same as the current asteroid belt \cite{Bottke05}. The orbits of these asteroids are taken from N-body simulations \cite{Morbidelli18} which were run to find the bombardment history of early Earth from rapidly depleted objects. The total initial mass of rapidly depleted asteroids was chosen using an iterative process in order to match the observation that 0.005~$M_\oplus$ \cite{Walker09} of highly siderophile elements (HSEs) should have been accreted to Earth during the Late Veneer from rapidly depleted asteroids. The number of rapidly depleted asteroids present in the inner Solar system decreases roughly exponentially with time due to a combination of ejection from the system by Jupiter, accretion onto Earth, and collisional evolution. The N-body simulations give the dynamical depletion of bodies with time, while our kinetic model gives the collisional depletion. By calculating the mass of asteroids accreted onto Earth when taking into account both the dynamical and collisional depletion, we found that an initial mass of $0.065~\mathrm{M}_\oplus$ of asteroids leftover after terrestrial planet formation led to $0.005~\mathrm{M}_\oplus$ of asteroids being accreted by Earth. We initialised this mass of asteroids in the kinetic model on the orbits from the N-body simulations. The model then accounted for the collisional depletion of asteroids, which produced smaller bodies and thus dust, and the dynamical depletion of asteroids, which removed bodies from the simulation according to the evolution found by the N-body results. The model therefore gave the distribution of dust produced by the collisional evolution of rapidly depleted asteroids leftover from planet formation. 

Having found the distribution of dust in the inner Solar system with time using the kinetic model, we then found the accretion rates of dust from each source population onto the early Earth as a function of time. The rate of each orbit overlapping with the Earth was found with the method of \cite{Wyatt10}, which was multiplied by the population of dust on each orbit to give the cosmic dust flux to Earth with time from each source.

\backmatter

\begin{itemize}
 \item [Data availability] All relevant data needed to evaluate the findings of the manuscript are included in the figures or in supplementary electronic materials.
 \item [Code availability] Details of specific codes used to perform simulations are available upon request.
 \item [Acknowledgments] C.R.W. acknowledges NERC and UKRI for support through a NERC DTP studentship, grant number NE/L002507/1; financial support from the Cambridge Leverhulme center for Life in the Universe; funding support from Trinity College (Cambridge) in the form of a Junior Research Fellowship; funding support from ETH Zürich and the NOMIS formation in the form of a research fellowship. Dr Dougal Ritson is thanked for their comments on an early version of the manuscript.
 \item[Contributions] C.R.W. conceived of the project, performed meta-data curation and analysis, and geological modelling. J.K.R. wrote the numerical model and performed all astrophysical simulations. O.S. and M.W. supervised the project. All authors contributed to the writing and editing of the manuscript text and figures.
 \item[Competing Interests] The authors declare no competing interests.
\end{itemize}

\includegraphics[width=13cm]{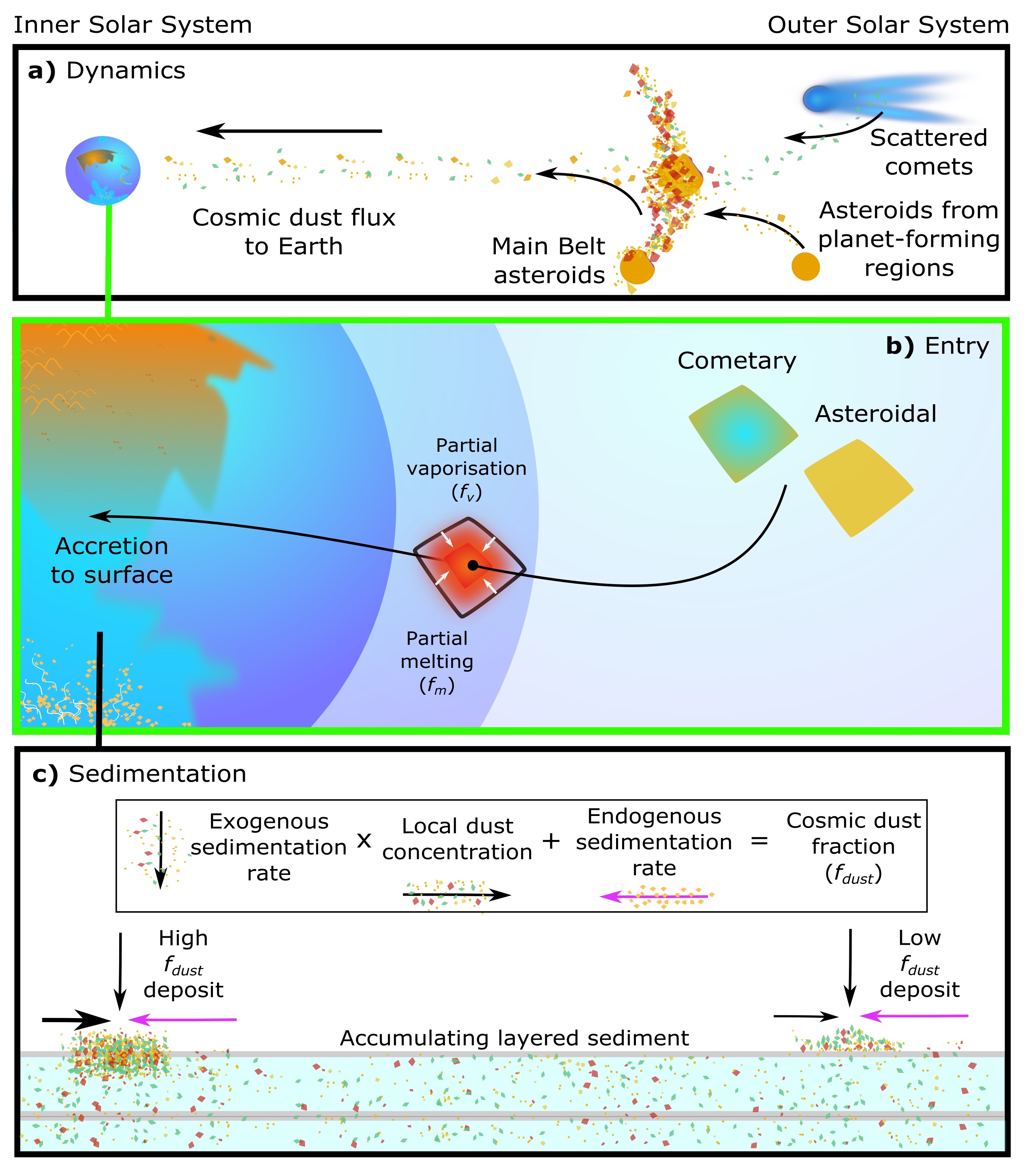}

\textbf{Figure 1: Delivery dynamics, atmospheric entry, and terrestrial sedimentation of cosmic dust.} A schematic illustration of factors considered in this study related to the formation of terrestrial sediments rich in cosmic dust. a) Dynamical sources of cosmic dust grains. Comets (undifferentiated) scattered inward from the Outer Solar System disintegrate to produce dust particles. Asteroids predominantly generate dust through collisions. b) Atmospheric entry of cosmic dust involves partial melting ($f_m$) and partial vaporisation ($f_v$), both of which influence the final mass of cosmic dust per unit mass of total sediment in a given sedimentary environment ($f_{dust}$, $\mathrm{kg/kg}$). c) The relative abundance of cosmic dust within terrestrial sediments is set by the local sedimentation rates of cosmic dust versus terrestrial (endogeneous) sediment, with dust proportion therefore being maximised in areas of low endogeneous sediment production and the action of local sedimentary concentration mechanisms.

\hspace{-1.8cm}\includegraphics[width=14cm]{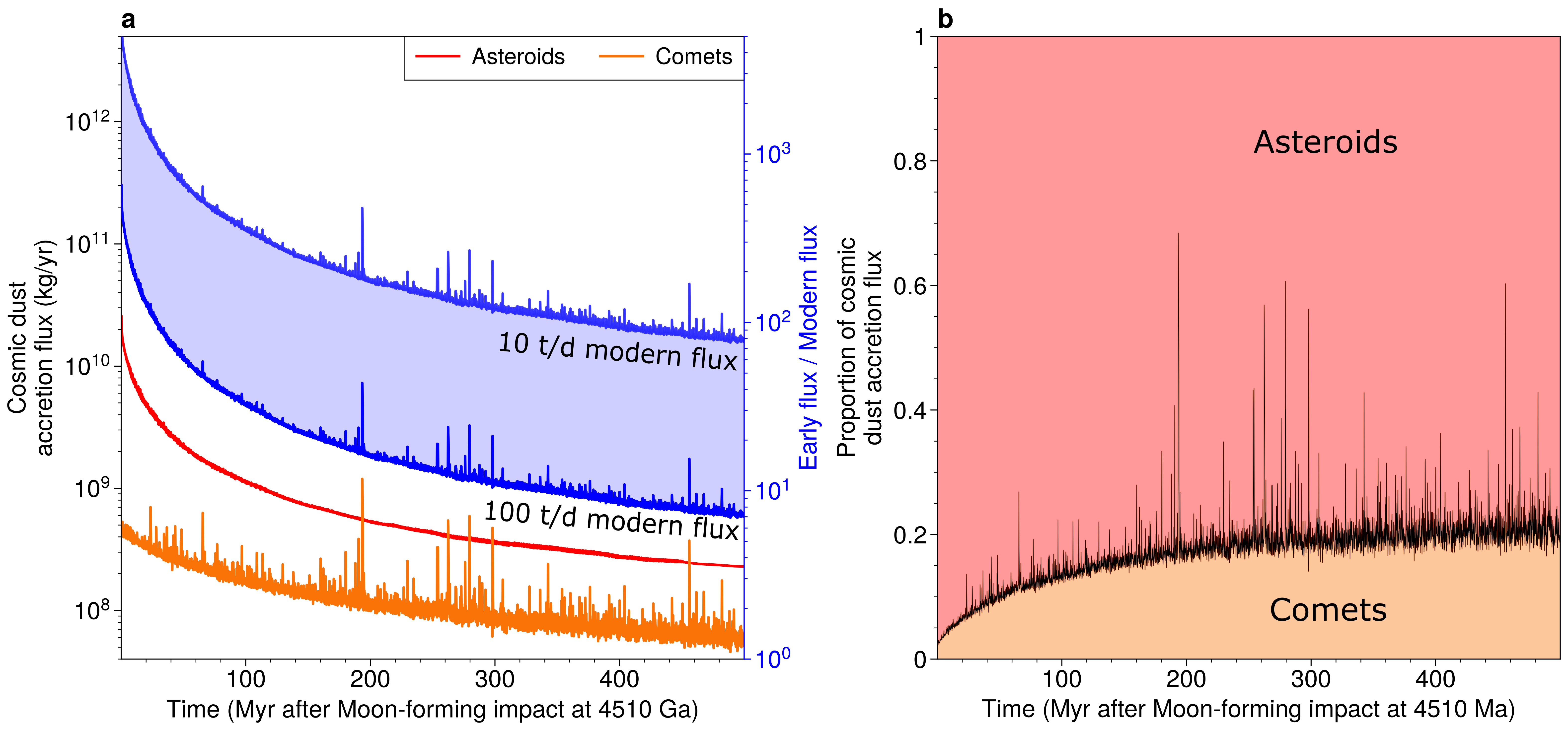}

\textbf{Figure 2: Interplanetary dust particle accretion flux to early Earth.} a) Total flux and individual fluxes of different cosmic dust populations - divided on the basis of parent body type. b) Proportions of the total cosmic dust accretion flux represented by each population over time. Boundaries between the proportion of each dust source are demarcated with solid black lines.

\begin{center}
{\includegraphics[width=11cm]{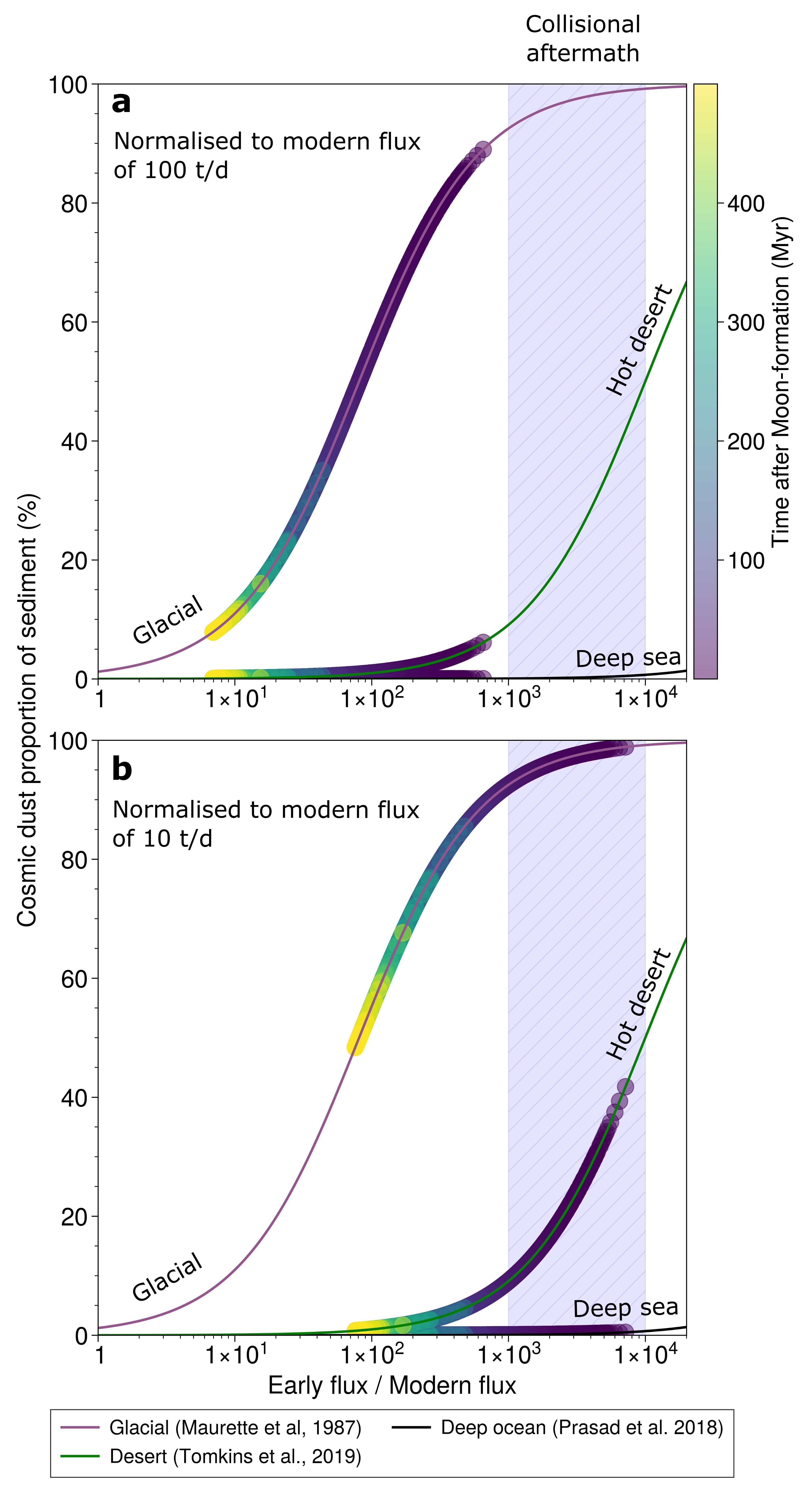}}
\end{center}

\textbf{Figure 3: Predicted proportion by mass of cosmic dust within sediments, which varies with the ratio of early to modern accreted dust flux.} {A conservatively high reference value for modern dust accretion is shown in (a). An optimistically low reference value is shown in (b).} The predicted proportion of dust within terrestrial sediments varies as a function of local environmental processes/conditions that preferentially concentrate dust relative to terrestrial sediment \cite{Maurette:1987, Prasad:2018, Tomkins:2019}. {Dust can be expected to represent a higher concentration of overall sediment in environments with low local endogenous sediment generation rates and where dust concentration mechanisms are active, e.g., sediment traps on glaciers, versus e.g., deep ocean sediments, where endogenous sedimentation rates are high and concentration mechanisms are ineffective (Fig. 1c). Individual data points, coloured according to their associated simulation time-step, indicate the average sediment compositions predicted by our simulations. The shaded region indicates the broad range of possible elevated dust fluxes that may transiently occur following individual large parent body collisions, which would translate into extremely cosmic-dust-rich sedimentary compositions in the context of our model.} 

\hspace{-1.5cm}\includegraphics[width=15cm]{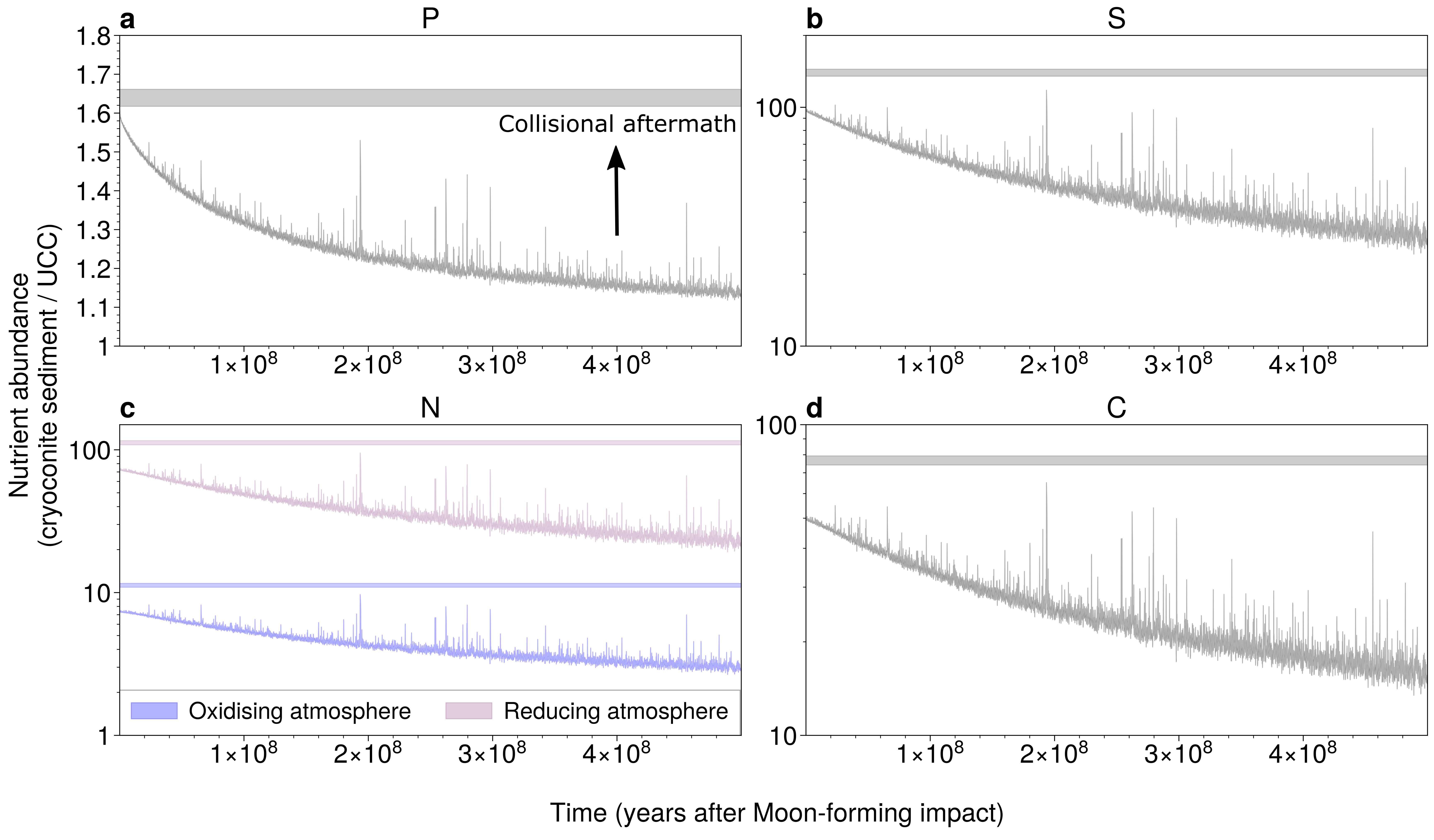}



{\textbf{Figure 4: a--d) Predicted chemical composition and utility for prebiotic chemistry of dust-rich sediments on early Earth.} Results are plotted for the absolute concentration of each element relative to the observed average concentration of emergent Upper Continental Crust on Earth, today} \cite{Rudnick:2014}. {Results plotted are obtained using a modern day dust flux of 10 t/d} \cite{Plane:2013}. {In c), both oxidising and reducing atmosphere scenarios are shown, due to the strong consequences for N-bearing organic molecule survival during entry under reducing conditions.}

\hspace{-1.5cm}\includegraphics[width=15cm]{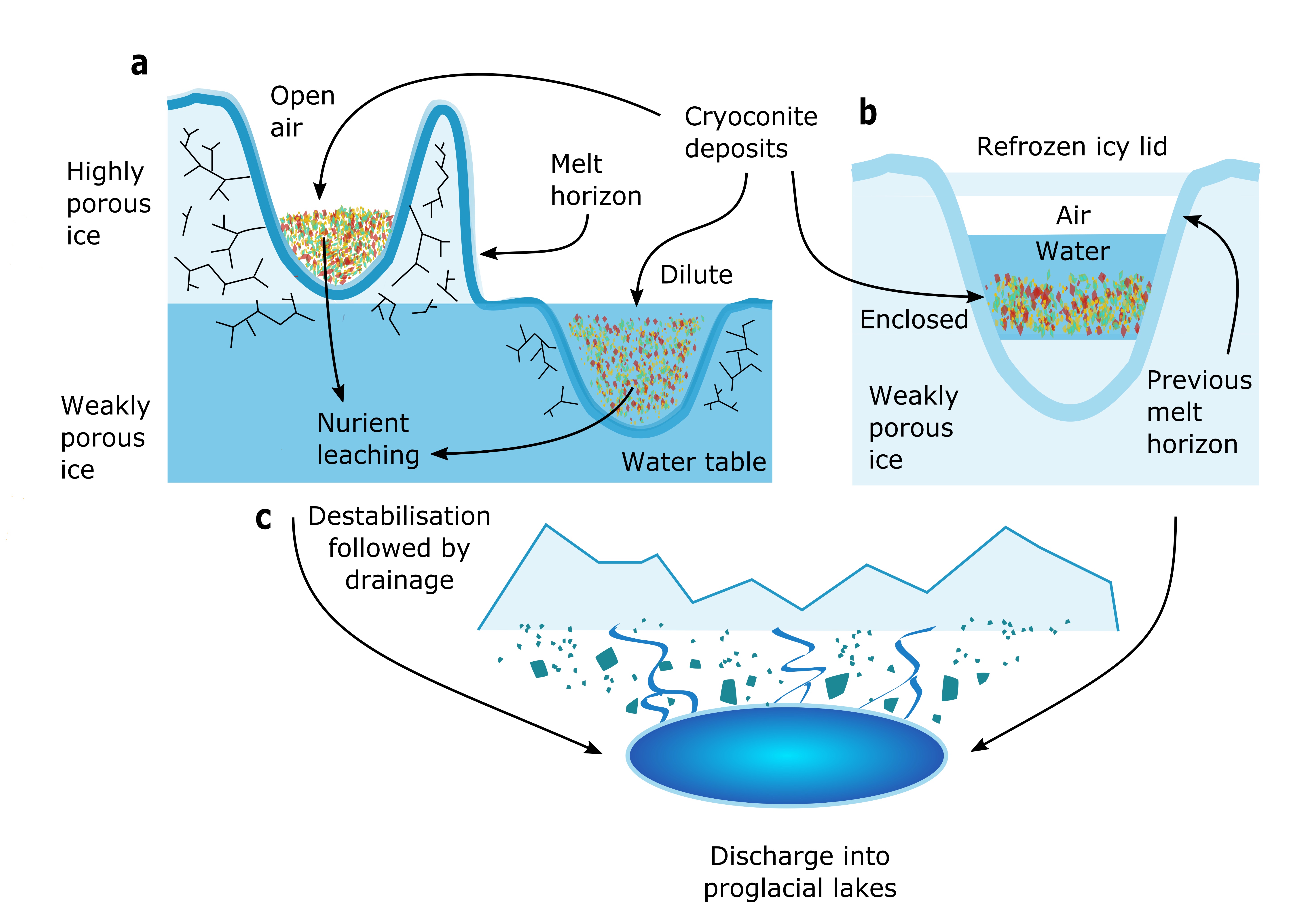}

{\textbf{Figure 5: Schematic illustration of cryoconite sedimentary deposit rich in cosmic dust.} a) Open air deposits may exist above the local water table, whereas dilute deposits lie in it. However, from a prebiotic chemistry perspective, both will suffer from nutrient leaching into a wider volume of diffuse water in the variably porous ice sheet. b) This is not the case for icy-lid cryoconite deposits, which are encased on all sides by weakly porous ice to cold impermeable ice. c) Regardless of their setting, cryoconite deposits are inherently unstable and most will be destabilised and drained within multi-annual timescales. Drained meltwater and cryoconite sediments will be transported in part to proglacial endorheic lakes, where longer term stockpiling of dust-derived species may occur.}



\newpage

\pagenumbering{gobble}

\begin{appendices}

\hspace{-1.2cm}\includegraphics[width=15cm]{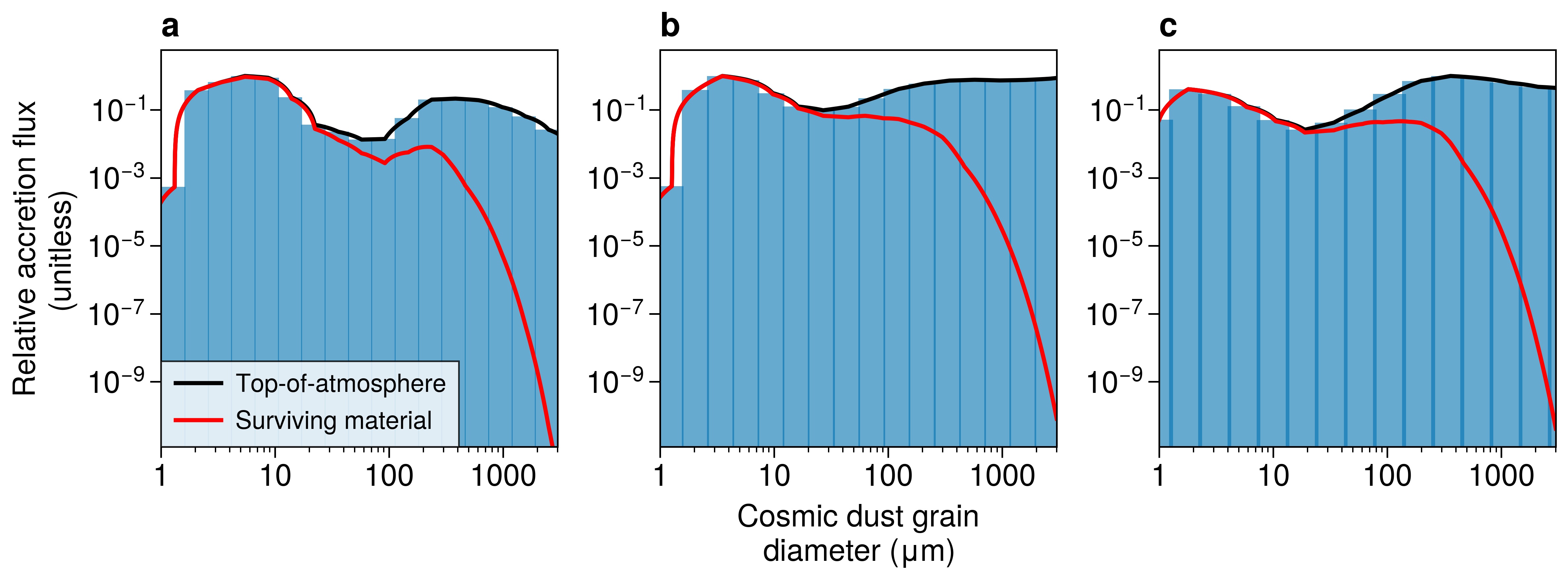}

\textbf{Supplementary Figure 1: Grain size distribution for the mean accretion flux of cosmic dust per type, before and after atmospheric entry.} Simulations predict the mean mass accreted for a given size bin, plotted as a histogram, and normalised in the case of each dust source to the maximum accretion flux among all bins. Smoothed distributions plot an interpolation between the centre-point of each bin. The bimodal distribution predicted by our numerical model for a) main belt asteroid, b) comet, and c) rapidly depleted asteroid sourced dust arriving at Earth's atmosphere undergoes a process of ablation that destroys larger grains with progressively greater efficiency. Results are unitless, being normalised to the highest accretion flux as a function of a grain size for each class of dust.

\newpage

\begin{tabular}{lllll}
Parameter & Symbol & Value (units) & Reference \\
\midrule
Nitrogen in cometary dust & $--$ & 2500 (ppm) & \cite{Matrajt:2003} \\ Carbon in cometary dust & $--$ & 50,000 (ppm) & \cite{Matrajt:2003} \\ Sulfur in cometary dust & $--$ & 90,400 (ppm) & \cite{Schramm:1990} \\ Phosphorus in cometary dust & $--$ & 1000 (ppm) & \cite{Wasson:1988} \\
\midrule
Nitrogen in asteroidal dust & $--$ & 1500 (ppm) & \cite{Wasson:1988} \\ Carbon in asteroidal dust & $--$ & 30,000 (ppm) & \cite{Wasson:1988} \\ Sulfur in asteroidal dust & $--$ & 60,000 (ppm) & \cite{Wasson:1988} \\ Phosphorus in asteroidal dust & $--$ & 1000 (ppm) & \cite{Walton:2021b} \\
\midrule
Nitrogen in upper continental crust & $--$ & 83 (ppm) & \cite{Rudnick:2014} \\ Carbon in upper continental crust & $--$ & 500 (ppm) & \cite{Javoy:1997} \\ Sulfur in upper continental crust & $--$ & 600 (ppm) & \cite{Rudnick:2014} \\ Phosphorus in upper continental crust & $--$ & 600 (ppm) & \cite{Rudnick:2014} \\
\midrule
Cosmic dust fraction in glacial trap & $\delta_{modern}$ & $1.52E-03$ & \cite{Suttle:2020} \\ Cosmic dust fraction in deep sea sediments & $\delta_{modern}$ & $7.00E-07$ & \cite{Prasad:2018} \\ Cosmic dust fraction in arid aeolian sediments & $\delta_{modern}$ & $8.00E-04$ & \cite{Tomkins:2019} \\
\midrule
Melted dust fraction in glacial trap & $f_m$ & $0.57$ & \cite{Suttle:2020} \\ Vaporised dust fraction & $f_v$ & $0.9$ & \cite{Suttle:2020} \\ Volatile loss fraction (N), unmelted & $f_{loss}$ & $0.9$ & \cite{Furi:2013, Marty:2005} \\ Volatile loss fraction (N), melted & $f_{loss}$ & $0.99$ & \cite{Furi:2013, Marty:2005} \\
\bottomrule
\end{tabular}

\textbf{Supplementary Table 1: Parameters used in estimating cosmic dust deposit compositions.} All compositional values for meteoritic components are model values, i.e., chosen to be representative of a group, rather than fully quantitative average compositions. Such an assessment would require compiling lines of evidence from different techniques, classes of meteoritic material, and forms of samples. Determining unbiased average compositions is an ongoing challenge in cosmochemistry and is beyond the scope of the present study. Noting this, we have chosen indicative and conservative values for parameters, on the basis of published literature.

\newpage

\includegraphics[width=0.8\linewidth]{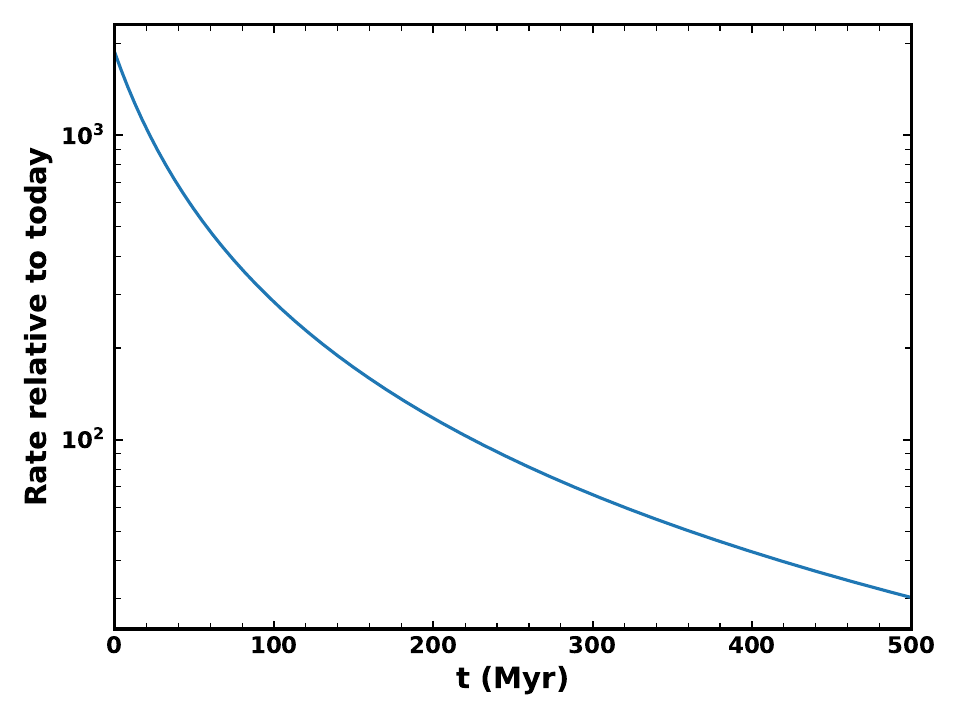}

\textbf{Supplementary Figure 2: The rate at which comets are input into the inner Solar system throughout the simulations. Data is plotted relative to the rate at which JFCs are scattered inward in the present Solar system}. Results from our simulations are plotted for the first 500 Myr of Solar System history.

\newpage

\includegraphics[width=0.8\linewidth]{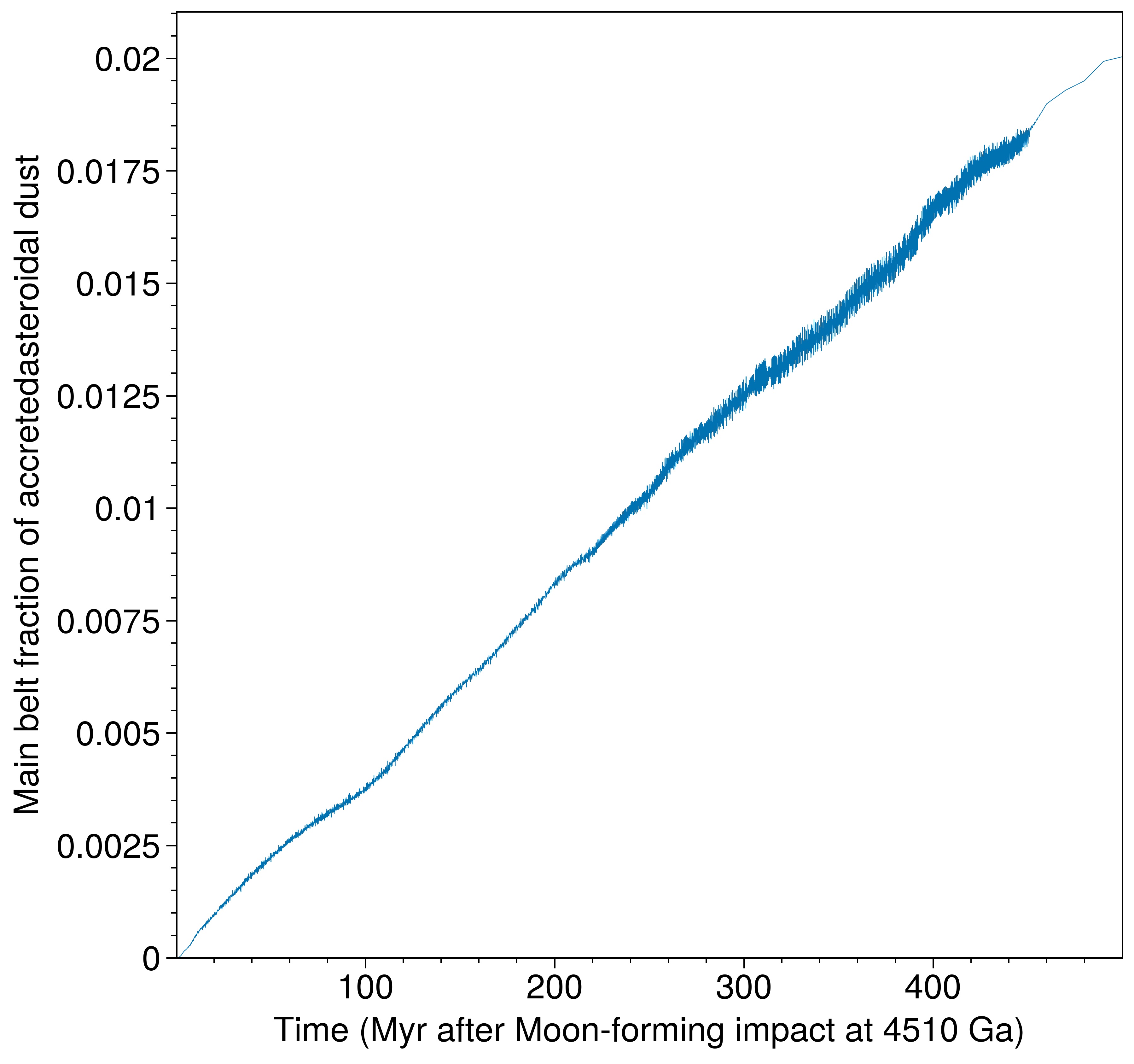}

\textbf{Supplementary Figure 3: Main belt fraction of asteroidal dust accreted to early Earth.} Results from our simulations are plotted for the first 500 Myr of Solar System history.

\end{appendices}

\end{document}